\documentstyle[12pt]{article}
\def\be{\begin{equation}}
\def\ee{\end{equation}}
\def\bea{\begin{eqnarray}}
\def\eea{\end{eqnarray}}

\begin{document}
\pagestyle{empty}
\begin{flushright}
{BROWN-HET-1014} \\
{September 1995}
\end{flushright}
\vspace*{5mm}
\begin{center}
{\bf The Classical $r$-Matrix for the Relativistic Ruijsenaars-Schneider
System}\\
[10mm]
by\\
[5mm]

J. AVAN\footnote{LPTHE Paris VI (CNRS-UA 280), Box 126, Univ. Paris 6, 4 Place
Jussieu, F-75252, Paris, France; and Brown University Physics Dept., Box 1843,
Brown Univ., Providence, RI  02912 USA} and G. ROLLET
\footnote{Northeastern University Physics Dept., Boston, MA  02115, USA.} \
\vskip 1.0in

{\bf ABSTRACT}
\end{center}
\vspace*{3mm}

\noindent We compute the classical $r$-matrix for the relativistic
generalization of the Calogero-Moser model, or Ruijsenaars-Schneider model, at
all values of the speed-of-light parameter $\lambda$.  We connect it with the
non-relativistic Calogero-Moser $r$-matrix $(\lambda \rightarrow -1)$ and the
$\lambda = 1$ sine-Gordon soliton limit.
\vfill
\newpage
\setcounter{page}{1}
\pagestyle{plain}
\renewcommand{\baselinestretch}{1.5}

\section{Introduction}

Our understanding of classical and quantum dynamical $r$-matrices has recently
known a number of interesting developments.  Dynamical classical $r$-matrices
were found for the Calogero-Moser models~\cite{1} and their spin
extension~\cite{2} in the rational, trigonometric/hyperbolic and elliptic
cases~\cite{3,4,5}.  This lead to the construction of observable algebras of
$W_n$-type~\cite{6} containing in particular the classical version of the
Yangian subalgebra already found as a symmetry of the spin (or Euler)
Calogero-Moser~\cite{7,8}.  Moreover a quantum analogue of the classical
(dynamical) Yang-Baxter equation for these $r$-matrices was found~\cite{9,10}
leading to the construction of commuting Hamiltonians for the spinless
(Calogero-Moser) and spin (Calogero-Sutherland) quantum systems~\cite{11}.

A relativistic-invariant extension of the Calogero-Moser model has been
constructed and shown to be integrable~\cite{12}.  It describes the dynamics of
solitons of the sine-Gordon model for a particular choice of
parameters~\cite{12,13}.  A spin extension was recently proposed~\cite{14}.

This model has several remarkable features.  First of all it was
shown~\cite{15} to be a Hamiltonian-type reduction of a simpler model
describing a free dynamics on a Heisenberg double~\cite{16} of a Lie group
(instead of a cotangent bundle as in the Calogero-Moser case).  This in turn
explained the duality symmetry of the corresponding quantum system~\cite{17}.
The trigonometric Ruijsenaars-Schneider model depends on two parameters,
corresponding to the speed-of-light and the radius of a compactifying circle.
In turn they are related to two parameters associated with the two copies of
the group Sl(N) which build the original Heisenberg double with its particular
Poisson structure.  Hence the RS model exhibits a self-duality property under
suitable interchange of these parameters.  Consequently also, the limits when
the speed-of-light parameter $\lambda$ goes to -1 (i.e. the non-relativistic
trigonometric CM model) and when the compactifying radius goes to $+ \infty$
(i.e. the rational relativistic RS model) are dual ( ``Ruijsennars
duality")~\cite{17} to each other.  Finally the wave functions of this model
were shown to be expressed by a path-integral formula for the G/G gauged
Wess-Zumino-Witten model~\cite{18}, a topological field theory over a cylinder
with a marked line.

Our main purpose in this paper will be to find the classical $r$-matrix of this
system for all values of its parameters, particularly $\lambda$.  Its existence
is guaranteed~\cite{19} by the explicit but cumbersome proof of commutation of
the Lax matrix adjoint invariants~\cite{12}.  Its structure should reflect the
Hamiltonian reduction procedure on the Heisenberg double.  Moreover the
Calogero-Moser $r$-matrix~\cite{3} arises from a non-relativistic limit, and
the ``Sine-Gordon Soliton" $r$-matrix~\cite{13} from a particular choice of the
speed-of-light parameter, making this $r$-matrix into an all-containing
structure for such models.

\section{The Ruijsenaars-Schneider Model}

The canonical variables are a set of rapidities $\{ \theta_i, i=1\cdots N\}$
and conjugate positions $q_i$ such that $\{ \theta_i, q_j\} = \delta_{ij}$.
The Hamiltonian is:
\be
H = mc^2 \sum_{j=1}^N \left( {\rm cosh} \, \theta_j \right) \prod_{k\not= j} \,
f (q_k - q_j)
\ee
where
\bea
f(q) & = & \left(1 + {g^2\over q^2} \right)^{1/2}\,\, ({\rm rational})\nonumber
\\
f(q) & = & \left( 1 + {\alpha^2\over {\rm sinh}^2 {\nu q\over
2}}\right)^{1/2}\,\,{\rm (hyperbolic)}\nonumber\\
&&\nonumber\\
f (q) & = & \left( \lambda + \nu {\cal P} (q) \right) \,\, {\rm ( elliptic)},
\,\,{\cal P} = {\rm Weierstrass\,\, function}
\eea
We shall be interested in the trigonometric/hyperbolic case.  The rational case
is obtained from an easy limit procedure as we shall soon see.  The elliptic
case is more involved.  Its integrability proof relies on specific identities
for elliptic functions~\cite{21} and we expect its $r$-matrix structure to be
very different from the $r$-matrices of the rational/trigonometric type.  Such
is the case in the non-relativistic limit (Calogero-Moser) for which the
Olshanetsky-Perelomov Lax matrix without spectral parameter~\cite{22} has yet
no known $r$-matrix and the usual ansatz~\cite{3} is know to be
inconsistent~\cite{20}.  This is ultimately due to the different nature of the
original dynamical system from which the elliptic model is
Hamiltonian-reduced~\cite{23}.

The Lax matrix reads:
\bea
&L =\sum_{j,k=1}^N \, L_{jk} \, e_{jk} \nonumber\\
&L_{jk} = \exp {\beta\over 2} \left( \theta_j + \theta_k \right) \cdot C_{jk}
\left( q_j - q_k \right) \cdot
\left( \displaystyle\prod_{m\not= j} \, f \left( q_j - q_m \right)
\prod_{l\not= k} \, f \left( q_l - q_k \right)\right)^{1/2}
\eea
where $\{ e_{jk} \}$ is the usual basis for $N \times N$ matrices; $f$ was
given in (2) and
\bea
C_{jk} (q) & = & {\gamma\over\gamma + iq} \,\,\,\,{\rm (rational)}\nonumber\\
C_{jk} (q) & = & \left( {\rm cosh} {\nu\over 2} q + ia \, {\rm sinh} {\nu\over
2} q \right)^{-1} \,\,{\rm (trigonometric)}
\eea
One has denoted in (4) $a = \sqrt{1-\alpha^{-2}},$ and $\gamma = \alpha/\nu$
for $\nu \rightarrow 0, \alpha\rightarrow 0$ in the rational limit.  The
Hamiltonian $H$ in (1) is $Tr \, L + Tr \, L^{-1}$ and the space-translation
generator $P$ is $Tr \, L - Tr\,L^{-1}$.  Connection with the more suitable
notations in~\cite{13} is obtained by introducing a ``Speed-of-light" parameter
$\lambda = {1 - ia\over 1+ia}$ and an exponentiated variable $z_i = e^{\nu
q_{i}}$.  One ends up with:
\bea
&L_{jk} = \exp {\beta\over 2} \left( \theta_j + \theta_k \right) \cdot \left(
z_j z_k \right)^{1/2} \cdot {\lambda+1\over 2}\nonumber\\
& \left[\displaystyle \prod_{l\not= j} \prod_{m\not= k} \, {(\lambda z_l + z_j
)^{1/2} (\lambda^{-1} z_l + z_j )^{1/2}\over z_l - z_j } \cdot {(\lambda
z_m+z_k)^{1/2} (\lambda^{-1} z_m+z_k)^{1/2}\over z_m-z_k} \right]^{1/2}
\eea

Note that when $\beta, \nu \rightarrow 0$, one can reabsorb $\sqrt{z_j z_k}$
into $\exp {\beta\over 2} (\theta_j + \theta_k)$ by a straightforward canonical
transformation.  Relativistic invariance is achieved by explicit introduction
of the speed of light $c$ and elimination of $\beta$.  As indicated in
{}~\cite{12} this requires a somewhat awkward redefinition of the canonical
variables $\theta_i$ and $z_i$ and we shall not use this parametrization here.
However it is important to know that $\alpha$ must be normalized as
${\alpha_0\over c}$ and $\nu$ as ${\nu_0\over c}$.  Hence the non-relativistic
limit $c\rightarrow \infty$ implies $\alpha \rightarrow 0$ and
$\lambda\rightarrow -1$; the ultrarelativistic limit $c\rightarrow 0$ implies
$\alpha\rightarrow + \infty $ and $\lambda = {1-i\over 1+i} $.  It is in this
sense that we sometimes call $\lambda$ the ``speed-of-light parameter" although
strictly speaking the compactification radius $\nu$ also contains $c$ as seen
here.  The since Gordon soliton Lax matrix~\cite{13} is obtained by setting
$\lambda = 1$.  The
non-relativistic limit is obtained by setting $\gamma \equiv g\beta$ (rational
case) and $\alpha = g\beta$ (trigonometric case) with $\beta \rightarrow 0$,
hence $\lambda \rightarrow - 1$.  One then gets.
\be
L_{jk} = \delta_{jk} + \beta \left\{ \theta_k \delta_{jk} + g {\sqrt{z_j
z_k}\over z_j - z_k} \right\} + 0 (\beta^2 )
\ee
and the order $\beta$-term is precisely the Calogero-Moser trigonometric Lax
matrix.  Note finally that the rational limit of (5) is obtained by setting
$\alpha \equiv \gamma \nu$ and $\nu = 0$.

\section{The Classical $r$-matrix}

We now solve the $r$-matrix equation for the trigonometric Lax operator (3).
The generic $r$-matrix structure is (for a Lax matrix $L$ in a Lie algebra $g$
):
\bea
\left\{ L \stackrel{\otimes}{,}L\right\} & \equiv & \sum_{ijkl=1}^N \, \left\{
L_{ij} , L_{kl} \right\} e_{ij}\otimes e_{kl} \ \in g \otimes g \nonumber\\
&=& \left[ r , L \otimes {\bf 1}\right]_{g\otimes g} - \left[ r^{\pi} , {\bf 1}
\otimes L\right]_{g\otimes g}
\eea
\be
r \equiv \sum_{ijkl=1}^n \, r_{ijkl} \, e_{ ij} \otimes e_{kl} \,\, ; \,\,
r^{\pi} = \sum \, r_{ijkl} \, e_{kl}\otimes e_{ij}
\ee
The Poisson brackets of our Lax matrix (5) read:
\bea
&&\left\{ L_{ij} , L_{kl} \right\} = {\nu\beta\over 8} L_{ij} L_{kl} \left\{
S_{il} \left( 1 - \delta_{ il }\right) + S_{ik} \left( 1 - \delta_{ik} \right)
+ S_{jl} \left( 1- \delta_{jl} \right)\right.\nonumber\\
&&+ S_{jk} \left( 1 - \delta_{jk} \right) + 2 \left(\left( G_{ij} -
G_{kl}\right)\right) \delta_{ik} + \left( G_{ij} + G_{kl} \right) \delta_{il}
\nonumber\\
&& \left.
+ \left( - G_{ij} - G_{kl} \right) \delta_{jk} + \left( - G_{ij} + G_{kl}
\right) \, \delta_{jl }\right\}
\eea
where:
\bea
G_{ij} & = & {\lambda z_i - z_j\over \lambda z_i + z_j } \quad F_{ij} \equiv
{z_i z_j\over z_i - z_j} \, \left( 1 - \delta_{ij}\right)\nonumber\\
S_{ij} & = & G_{ij} - G_{ji} - 2 F_{ij}
\eea
The Poisson bracket structure (9) has a number of features which will be
helpful in finding the $r$-matrix.  First of all, it exhibits a quadratic
behavior in the Lax matrix, dressed by the quantities $S $ and $G$.  This leads
us to conjecture for the $r$-matrix a linear behavior, similarly dressed by
$z$-dependent functions.

Then the indices carried by the Lax elements $L$ and the dressing functions $S,
G$ on the r.h.s. of (8) are only the original indices $i,j,k,l$ and no extra
index ever occurs.  Moreover the dressing functions $S$ and $G$ are rational
functions of the sole $z$ variable of which they carry the index.  This
``locality" property of (9) leads us, in a first step towards solving (7), to
set restrictions on the algebraic structure of $r$ in (8) from the following
argument:

The $r$-matrix structure (7) generates a priori a Poisson bracket $\{ L_{ij} ,
L_{kl}\} $ containing a summation over one extra ``free" index from the
products $r\cdot L\otimes {\bf 1}$ and $L\otimes {\bf 1} \cdot r \cdots$
However as we have just seen (9) does not exhibit such a summation.

This in particular strongly precludes the existence in $r$ of terms with four
distinct indices.  In fact, these considerations may be extended to lower index
terms, leading to the Ansatz:
\bea
r_{ijkl}& =& {\nu\beta\over 8} \left\{ A_{ikl} \, L_{kl} \, \delta_{ij} +
B_{ij} \left( L_{jl} \, \delta_{jk} + L_{kj} \, \delta_{il}\right)\right.
\nonumber\\
&+& \left. C_{ij} \left( L_{ki} \, \delta_{jl} + L_{jl} \, \delta_{ik}
\right)\right\}
\eea
when $A,B$ and $C$ are furthermore assumed (as a consequence of the similar
property in (9)) to be rational functions of the sole dynamical variables $\{
z_i\} $ of which they carry the explicit index.  The quadratic behavior of (9)
in $L$ is also taken into account by this linear-in-$L$ ansatz.  The second
step of our reasoning takes advantage of the pole structure of (9):

Plugging (11) into (7) and comparing it with the explicit expression (9) at the
particular points $\lambda z_i = - z_j$ leads us to setting:
\be
A_{ikl} = {1\over 2} \left( S_{il} + S_{ki} \right) + 2 G_{il}\left( 1 -
\delta_{ik} \right) - 2 G_{ki} \left( 1 - \delta_{il} \right)
\ee
It finally turns out from a careful inspection of the remaining terms that the
Poisson structure (9) is fully reproduced by (11) provided one also sets:\be
B_{ij} = - 2 F_{ij} \,\,; \,\, C_{ij} = 0
\ee

The explicit checking of the consistency of this form for the $r$-matrix is
considerably simplified by using a number of functional identities connecting
$L, F, G$ and allowing some permutations of indices in the quadratic
expressions $L_{ij} L_{kl}$.  For instance one has:
\be
\left( F_{ik} + F_{jl} \right) L_{il} L_{kj} = \left( F_{ik} + F_{jl} + \left(
G_{kj} - G_{il} \right) \left( 1 - \delta_{ik} - \delta_{jl} \right)\right)
L_{ij} L_{kl}
\ee

\noindent We interprete both the ``quadratic" form of $r$ in (11) and the role
played by such functional equations as (14) in checking the $r$-matrix
structure as a reflection of the Hamiltonian reduction procedure from a
Heisenberg double.  The quadratic $r$-matrix, in particular, is characteristic
of such integrable systems and the ``dressing" terms $A, B, C$ may viewed as
generated by the Hamiltonian reduction, in the same way as the dynamical terms
in the $r$-matrix of Calogero-Moser were generated by the Marsden-Weinstein
reduction procedure~\cite{5}.  Clarifying these issues will be left for further
studies.

\section{Limits of the $r$-Matrix:  Calogero-Moser and Sine Gordon}

The non-relativistic limit of the Lax matrix (3) yielding the trigonometric
Calogero-Moser model was described in the introduction (6).  Using the same
reparametrization $\alpha = g\beta, \beta\rightarrow 0$, the $r$-matrix (11)
becomes:
\bea
r_{ijkl} & = & - \nu \beta \left\{ {1\over 4} \, \delta_{ij} \left( \delta_{il}
+ \delta_{ik} \right) \left( 1 - \delta_{kl}\right) {1\over {\rm sinh}
{\nu\over 2} (q_k-q_l)} \right.\nonumber\\
&& +\left. {1\over 2} \left( 1 - \delta_{ij} \right) \delta_{il} \delta_{jk}
{\rm cotanh} {\nu\over 2} \left( q_i - q_j \right) \right\} + 0 (\beta^2 )
\eea
which is precisely the well-known trigonometric C.M. matrix~\cite{3}.

The sine Gordon limit is not immediately identifiable with the $r$-matrix
obtained in~\cite{13} which is given by setting $A = 0, B=C=F$ in (11).  Of
course an $r$-matrix is not unique and has actually a large ``moduli space".
In this case, the difference $\tilde{r}$ between the RS matrix (11-13) and the
particular sine-Gordon soliton $r$-matrix in~\cite{13} obeys:
\bea
&\left[ \tilde{r}, L (\lambda = 1) \otimes {\bf 1})\right] - \left[
\tilde{r}^{\pi} , {\bf 1} \otimes L \right] = 0 \nonumber\\
&{\rm but} \,\, \left[ \tilde{r} , L\otimes {\bf 1} \right] \not= 0
\eea

At this moment we do not know how to generalize $\tilde{r}$ to $(\lambda\not=
1)$.  In fact $\lambda = 1$  is the particular point where $L$ is symmetric,
and the rational fractions $S, G$ and $F$ simplify dramatically.  It seems
actually that the 3-index ansatz (11) is the only generic one  and the
reduction to 2-index functions is particular to the symmetric point $\lambda =
1$.

\section*{Acknowledgements}

This work was sponsored by Lavoisier Grant, French Foreign Affair Ministry,
Paris, France (G.R.); CNRS; CNRS-NSF Exchange Programme AI-06-93;  and DOE
Grant DE-FG02-91ER40688, Task A, Brown University. G. R. wishes to thank LPTHE
Paris VI for its hospitality.  J. A. thanks O. Babelon for discussions, and
Brown University Physics Department for their kind support.

\end{document}